\documentclass[preprint,12pt]{elsarticle}

\usepackage{dcolumn}

\usepackage{graphicx}

\biboptions{sort&compress}

\begin{document}

\begin{frontmatter}

\title{On the Hellmann-Feynman theorem for statistical averages}

\author{Francisco M. Fern\'{a}ndez}

\address{INIFTA (UNLP, CCT La Plata-CONICET), Divisi\'{o}n Qu\'{i}mica Te\'{o}rica,\\
Blvd. 113 y 64 (S/N), Sucursal 4, Casilla de Correo 16,\\
1900 La Plata, Argentina}

\ead{fernande@quimica.unlp.edu.ar}

\begin{abstract}

We discuss the Hellmann-Feynman theorem for degenerate states and
its application to the calculation of the derivatives of
statistical averages with respect to external parameters.

\end{abstract}

\end{frontmatter}

Some time ago there was a discussion about the validity of the
Hellmann-Feynman theorem (HFT)\cite{F39} for degenerate
states\cite{ZG02,F04,V04,BHM04}. Recently, some of those
results\cite{F04,V04,BHM04} proved useful in deriving an
expression for the derivative of the non-extensive free energy
with respect to an external parameter\cite{R13}. The author took
into account the possible occurrence of degenerate states in the
proof of his Lema 1\cite{R13}. However, in the proof of his
Theorem 2 he appears to assume that the eigenvalues of $\hat{H}$
and of the observable $\hat{A}$ are nondegenerate\cite{R13}. We
think that this discrepancy should be analyzed carefully. In this
letter we investigate the connection between the HFT for
quantum-mechanical expectation values and statistical averages
when there are degenerate states.

The starting point of our discussion is the Schr\"{o}dinger equation
\begin{equation}
\hat{H}\psi _{m}=E_{m}\psi _{m}  \label{eq:Schro}
\end{equation}
where $m$ is a set of quantum numbers that completely specify the stationary
state $\psi _{m}$ and we assume that $\left\langle \psi _{n}\right| \left.
\psi _{m}\right\rangle $ $=\delta _{mn}$. If the Hamiltonian operator $%
\hat{H}$ depends on a parameter $\lambda $ then its eigenvalues and
eigenvectors will also depend on it. Following Rastegin\cite{R13} we assume
that the spectrum of $\hat{H}$ is discrete.

The Hellmann-Feynman theorem for nondegenerate states does not present any
difficulty and for this reason we assume that the energy level $E_{n}$ is $%
g_{n}$-fold degenerate:
\begin{equation}
\hat{H}\psi _{ni}=E_{ni}\psi _{ni},\;E_{ni}=E_{n},\;i=1,2,\ldots ,g_{n}
\label{eq:Schro_deg}
\end{equation}
If we differentiate this equation with respect to $\lambda $ and then apply
the bra $\left\langle \psi _{nj}\right| $ from the left, we obtain
\begin{equation}
\left\langle \psi _{nj}\right| \frac{\partial \hat{H}}{\partial \lambda }%
\left| \psi _{ni}\right\rangle =\frac{\partial E_{ni}}{\partial \lambda }%
\delta _{ij}  \label{eq:HFT_deg_1}
\end{equation}
This equation tells us that there is a set of degenerate
eigenvectors for which the diagonal HFT ($i=j$) is always valid.
For simplicity we avoid a detailed discussion of the
differentiation of eigenvectors and operators with respect to the
external parameter; in this respect we follow earlier approaches
to the subject\cite{F04,V04,BHM04}.

It is convenient to analyse two different cases separately. The simpler one
takes place when $g_{n}$ does not change with $\lambda $ (at least for all
values of physical interest of this external parameter). Any unitary
transformation of the degenerate states
\begin{equation}
\chi _{i}=\sum_{j=1}^{g_{n}}c_{ji}\psi _{nj},\;i=1,2,\ldots ,g_{n}
\label{eq:unit_trans}
\end{equation}
yields a set of $g_{n}$ eigenvectors of $\hat{H}$ with eigenvalue $E_{n}$.
They satisfy
\begin{equation}
\left\langle \chi _{i}\right| \frac{\partial \hat{H}}{\partial \lambda }%
\left| \chi _{j}\right\rangle =\sum_{k=1}^{g_{n}}c_{ki}^{*}c_{kj}\frac{%
\partial E_{nk}}{\partial \lambda }  \label{eq:HFT_deg_2}
\end{equation}
Since $g_{n}$ does not change with $\lambda $ it is obvious that $\frac{%
\partial E_{nk}}{\partial \lambda }=\frac{\partial E_{n}}{\partial \lambda }$
for all $k=1,2,\ldots ,g_{n}$ and this equation simplifies to
\begin{equation}
\left\langle \chi _{i}\right| \frac{\partial \hat{H}}{\partial \lambda }%
\left| \chi _{j}\right\rangle =\frac{\partial E_{ni}}{\partial \lambda }%
\delta _{ij}  \label{eq:HFT_deg_3}
\end{equation}
that is similar to (\ref{eq:HFT_deg_1}). In other words: in this
case we do not have to worry about choosing a particular set of
eigenvectors and all the results derived by Rastegin\cite{R13}
apply to any observable provided that degeneracy is not removed
through variations of $\lambda$.

When $g_{n}$ changes, for example at $\lambda =\lambda _{0}$, then
$\left. \frac{\partial E_{ni}}{\partial \lambda }\right| _{\lambda
=\lambda _{0}} \neq \left. \frac{\partial E_{nj}}{\partial \lambda
}\right| _{\lambda =\lambda _{0}}$ for some $i\neq j$ and
Eq.~(\ref{eq:HFT_deg_3}) does not follow from
Eq.~(\ref{eq:HFT_deg_2}). However, in this case we can derive the
equation\cite{F04}
\begin{equation}
\sum_{i=1}^{g_{n}}\left\langle \chi _{i}\right| \frac{\partial \hat{H}}{%
\partial \lambda }\left| \chi _{i}\right\rangle =\sum_{k=1}^{g_{n}}\frac{%
\partial E_{nk}}{\partial \lambda }  \label{eq:HFT_sum}
\end{equation}
that was invoked by Rastegin\cite{R13} to prove his Lemma 1. Typically, $%
g_{n}(\lambda )<g_{n}(\lambda _{0})$ which happens, for example, when the
symmetry of the system is greater when $\lambda =\lambda _{0}$.

Before discussing the trace averages that currently appear in statistical
mechanics, it is convenient to analyse this problem from another point of
view. If we differentiate Eq.~(\ref{eq:Schro}) with respect to $\lambda $
and then apply the bra $\left\langle \psi _{n}\right| $ from the left, we
obtain an expression for both the diagonal ($m=n$) and off-diagonal ($m\neq n
$) HFT\cite{C63}
\begin{equation}
\left\langle \psi _{n}\right| \frac{\partial \hat{H}}{\partial \lambda }%
\left| \psi _{m}\right\rangle =\left( E_{m}-E_{n}\right) \left\langle \psi
_{n}\right| \left. \frac{\partial \psi _{m}}{\partial \lambda }\right\rangle
+\frac{\partial E_{m}}{\partial \lambda }\delta _{mn}
\label{eq:HFT_off-diag}
\end{equation}
If the eigenvalues $E_{m}$ and $E_{n}$ are degenerate at $\lambda =\lambda
_{0}$ $E_{m}(\lambda _{0})=E_{n}(\lambda _{0})$ then
\begin{equation}
\left. \left\langle \psi _{n}\right| \frac{\partial \hat{H}}{\partial
\lambda }\left| \psi _{m}\right\rangle \right| _{\lambda =\lambda
_{0}}=\left. \frac{\partial E_{m}}{\partial \lambda }\right| _{\lambda
=\lambda _{0}}\delta _{mn}  \label{eq:HFT_deg}
\end{equation}
This equation is identical to Eq.~(\ref{eq:HFT_deg_1}) but its derivation
reveals that the diagonal HFT applies to degenerate states provided that we
choose the eigenvectors of $\hat{H}$ according to
\begin{equation}
\psi _{n}(\lambda _{0})=\lim\limits_{\lambda \rightarrow \lambda _{0}}\psi
_{n}(\lambda )  \label{eq:lim_psi}
\end{equation}

In what follows we analyse the HFT in the context of statistical-averages
that we develop in a somewhat more general setting than that considered by
Rastegin\cite{R13}. For any Hermitian operator $\hat{W}$ that commutes with $%
\hat{H}$:
\begin{equation}
\left[ \hat{H},\hat{W}\right] =0  \label{eq:[H,W]=0}
\end{equation}
the hypervirial theorem
\begin{equation}
\left\langle \psi _{i}\right| \left[ \hat{H},\hat{W}\right] \left| \psi
_{j}\right\rangle =\left( E_{i}-E_{j}\right) \left\langle \psi _{i}\right|
\hat{W}\left| \psi _{j}\right\rangle   \label{eq:hypervir}
\end{equation}
tells us that
\begin{equation}
\left\langle \psi _{i}\right| \hat{W}\left| \psi _{j}\right\rangle =0\mathrm{%
\ }\ \mathrm{if}\ E_{i}\neq E_{j}  \label{eq:Wij=0}
\end{equation}

If the trace
\begin{equation}
tr\left( \hat{W}\frac{\partial \hat{H}}{\partial \lambda }\right)
=\sum_{i}\sum_{j}\left\langle \varphi _{i}\right| \hat{W}\left| \varphi
_{j}\right\rangle \left\langle \varphi _{j}\right| \frac{\partial \hat{H}}{%
\partial \lambda }\left| \varphi _{i}\right\rangle   \label{eq:tr(WdH)_gen}
\end{equation}
exists then it is invariant under unitary transformations of the basis set
and we can thus choose the eigenvectors of $\hat{H}$ (\ref{eq:lim_psi}) that
satisfy Eq.~(\ref{eq:HFT_deg}). Since they also satisfy Eq.~(\ref{eq:Wij=0})
we have
\begin{equation}
tr\left( \hat{W}\frac{\partial \hat{H}}{\partial \lambda }\right)
=\sum_{n}\left\langle \psi _{n}\right| \hat{W}\left| \psi _{n}\right\rangle
\left\langle \psi _{n}\right| \frac{\partial \hat{H}}{\partial \lambda }%
\left| \psi _{n}\right\rangle =\sum_{n}\left\langle \psi _{n}\right| \hat{W}%
\left| \psi _{n}\right\rangle \frac{\partial E_{n}}{\partial \lambda }
\label{eq:tr(WdH)}
\end{equation}
This expression is the basis for many of the results derived by
Rastegin\cite{R13} such as, for example, his Lemma 1:
\begin{equation}
tr\left[ f(\hat{H})\frac{\partial \hat{H}}{\partial \lambda }\right]
=\sum_{n}f(E_{n})\frac{\partial E_{n}}{\partial \lambda }
\label{eq:tr(F(H)dH)}
\end{equation}
We appreciate that we do not have to worry about degeneracy when calculating
traces provided that $\hat{W}$ is diagonal with respect to the nondegenerate
eigenvectors of $\hat{H}$. In other words: we do not need to invoke the HFT
sum expression (\ref{eq:HFT_sum}).

Equation (\ref{eq:tr(WdH)}) also applies to any operator $\hat{A}$ that
depends on a parameter $\lambda $, exhibits a discrete spectrum and commutes
with $\hat{W}$. Following Rastegin\cite{R13} we choose an element of the
complete set of commuting observables that shares a common eigenbasis with $%
\hat{H}$%
\begin{equation}
\hat{A}\psi _{m}=a_{m}\psi _{m}  \label{eq:A_psi}
\end{equation}
In such a case we have
\begin{equation}
tr\left( \hat{W}\frac{\partial \hat{A}}{\partial \lambda }\right)
=\sum_{n}\left\langle \psi _{n}\right| \hat{W}\left| \psi _{n}\right\rangle
\frac{\partial a_{n}}{\partial \lambda }  \label{eq:tr(WdA)}
\end{equation}
provided that
\begin{equation}
\left. \left\langle \psi _{n}\right| \frac{\partial \hat{A}}{\partial
\lambda }\left| \psi _{m}\right\rangle \right| _{\lambda =\lambda
_{0}}=\left. \frac{\partial a_{m}}{\partial \lambda }\right| _{\lambda
=\lambda _{0}}\delta _{mn}  \label{eq:HFT_deg_A}
\end{equation}
when $E_{m}(\lambda _{0})=E_{n}(\lambda _{0})$ and $E_{m}(\lambda )\neq
E_{n}(\lambda )$ for $\lambda \neq \lambda _{0}$.

Rastegin's equations (27) and (31) that are necessary for proving his
theorem 2\cite{R13} require that the eigenvectors satisfy present equations (%
\ref{eq:HFT_deg}) and (\ref{eq:HFT_deg_A}) when $g_{n}$ changes at $\lambda
_{0}$. If one does not state these conditions explicitly then one is in
principle assuming that $g_{n}$ does not change with $\lambda $ and the
resulting theorems are not so widely applicable.

\end{document}